\documentclass{article}

\usepackage{PRIMEarxiv}

\usepackage[utf8]{inputenc} 
\usepackage[T1]{fontenc}    
\usepackage{hyperref}       
\usepackage{url}            
\usepackage{booktabs}       
\usepackage{amsfonts}       
\usepackage{nicefrac}       
\usepackage{microtype}      
\usepackage{lipsum}
\usepackage{fancyhdr}       
\usepackage{graphicx}       

\usepackage{color}
\usepackage{amsmath}
\usepackage{amssymb}
\usepackage{setspace}

\usepackage{lineno}

\usepackage{bm}
\usepackage{multirow}

\graphicspath{{media/}}     

\title{Multiscale
Physics-Informed Neural Networks for the Inverse Design of Hyperuniform Optical Materials
}

\author{
  Roberto Riganti \\
  Physics Department \\
  Boston University,  Boston\\
   \And
  Yilin Zhu \\
  Division of Materials Science \& Engineering \\
  Boston University,  Boston\\
  \And
  Wei Cai \\
  Department of Mathematics, \\
  Southern Methodist University, Dallas
  \And
  Salvatore Torquato \\
  Department of Chemistry, Department of Physics, \\and Princeton Materials Institute \\
  Princeton University, Princeton\\
  \And
  Luca Dal Negro\textsuperscript{*} \\
  Department of Electrical \& Computer Engineering, \\
  Division of Materials Science \& Engineering,\\
  Physics Department, and Photonics Center \\
  Boston University, Boston\\
  \textsuperscript{*}\texttt{dalnegro@bu.edu} \\
}

\begin{document}
\maketitle

\begin{abstract}
In this article, we employ multiscale physics-informed neural networks (MscalePINNs) for the inverse design of finite-size photonic materials with stealthy hyperuniform (SHU) disordered geometries. Specifically, we show that MscalePINNs can capture the fast spatial variations of complex fields scattered by arrays of dielectric nanocylinders arranged according to isotropic SHU point patterns, thus enabling a systematic methodology to inversely retrieve their effective dielectric profiles. Our approach extends the recently developed high-frequency homogenization theory of hyperuniform media and retrieves more general permittivity profiles for applications-relevant finite-size SHU systems, unveiling unique features related to their isotropic nature. In particular, we numerically corroborate the existence of a transparency region beyond the long-wavelength approximation, enabling effective and isotropic homogenization even without disorder-averaging, in contrast to the case of uncorrelated Poisson random patterns. The flexible multiscale network approach introduced here enables the efficient inverse design of more general effective media and finite-size optical metamaterials with isotropic electromagnetic responses beyond the limitations of traditional homogenization theories.
\end{abstract}

\keywords{Stealthy Hyperuniform \and Neural Networks \and Homogenization \and Multiscale}

\section{\label{sec:intro}Introduction}
Disordered hyperuniform systems were recently discovered in a variety of contexts and phenomena, including glass formation, spin systems, photonic band structures and radiation engineering, nanophotonics, and biological systems, to name a few~\cite{torquato_local_2003, torquato_hyperuniform_2018}. A hyperuniform point pattern is characterized by the vanishing of its structure factor $S(\bm{k})$ when the wavevector goes to zero, resulting in the suppression of long-wavelength density fluctuations~\cite{torquato_ensemble_2015, klatt_wave_2022}. In the context of condensed matter physics, it was shown that two- and three-dimensional systems of particles can freeze into highly degenerate disordered hyperuniform states at zero temperature with stealthy hyperuniform (SHU) point pattern geometry, challenging the traditional belief that liquids freeze into highly symmetric structures~\cite{ma_random_2017, sgrignuoli_subdiffusive_2022,izrailev_anomalous_2012}. Among the hyperuniform states of matter, SHU systems are characterized by a structure factor that vanishes over a compact interval of wavevectors. Therefore, stealthy hyperuniformity is a stronger condition than standard hyperuniformity because single scattering events are prohibited within a large interval of spatial frequencies, thus suppressing the corresponding far-field radiation over sizeable angular ranges. Importantly, the structural correlation properties of disordered SHU media can be largely controlled by the $\chi$ stealthiness parameter, which equals the ratio of the number of constrained wave vectors in reciprocal space to the total number of degrees of freedom, providing opportunities for tuning the structures in between traditional (uncorrelated) random media for $\chi=0\%$ and highly-correlated (periodic) structures for $\chi=100\%$. Moreover, it was established that the degree of short-range order in these systems increases with $\chi$, inducing a transition from disordered to crystalline phases when $\chi>50\%$ in two spatial dimensions. 
Recently, the interaction of hyperuniform media with electromagnetic waves attracted significant interest resulting in the discovery of amorphous materials with large and complete photonic band gaps, photon sub-diffusion and localization, as well as in the engineering of enhanced light absorbers, quantum cascade lasers, directional extractors of incoherent emission for light-emitting diodes, free-form waveguides, and Luneburg lenses~\cite{chen_designing_2018, cheben_subwavelength_2018, gkantzounis_hyperuniform_2017, man_isotropic_2013, sgrignuoli_subdiffusive_2022, deglinnocenti_hyperuniform_2016, zhang_experimental_2019,gorsky_engineered_2019,sun_imaging_2018}. Moreover, the effective electromagnetic wave properties of stealthy hyperuniform systems have been studied beyond the quasistatic regime within a rigorously valid nonlocal theory in the thermodynamic limit of infinite system size, leading to the prediction of perfect transparency intervals up to finite wavenumbers~\cite{torquato_nonlocal_2021,kim_theoretical_2024,alhaitz_experimental_2023,aubry_experimental_2020}. This characteristic "transparency regime" is manifested by a vanishing imaginary part of the effective dielectric constant $\epsilon_e$ of the scattering structure within a prescribed range of wavelengths~\cite{torquato_nonlocal_2021,kim_theoretical_2024}. Recent work has shown that this transparency is robust against multiple scattering, opening the doors to numerous improvements in the design of photonic materials with applications ranging from light harvesting in solar cells \cite{vynck_photon_2012} to the design of waveguides \cite{milosevic_hyperuniform_2019, cheron_wave_2022, romero-garcia_wave_2021, kim_multifunctional_2020, gkantzounis_freeform_2017}. However, dynamic homogenization theory cannot be directly applied to finite-size structures when the strength of the multiple scattering renders the effective permittivity spatially dependent. This limitation establishes the need for a more general predictive approach intended for the inverse design of the effective wave characteristics of disordered hyperuniform media and photonic devices~\cite{zhang_reconfigurable_2022, florescu_designer_2009, shi_computational_2023}. 

In this article, we propose and develop an accurate and flexible deep-learning methodology to design novel optical materials by predicting the effective electromagnetic properties of finite-size hyperuniform structures in the dynamic scattering regime based on multiscale physics-informed neural networks (MscalePINNs). Specifically, we apply this approach to disordered SHU and Poisson arrays of dielectric nanopillars of radius $a$ with constant relative permittivity $\epsilon_r$ and demonstrate enhanced transparency enabling the accurate inverse retrieval of the effective dielectric permittivity $\hat{\epsilon}(x,y;k)$ of SHU structures with different sizes, shapes, and dielectric contrast values. Our results for SHU structures show accurate inverse retrieval of dielectric properties without averaging over multiple disorder realizations, in contrast to the case of the uncorrelated Poisson patterns of equivalent density. Moreover, we show that SHU arrays can be effectively homogenized at shorter wavelengths compared to Poisson arrays with identical particle volume fractions, conjecturing that SHU structures are transparent over a wider range of wavelengths even in finite-size systems. Importantly, we also establish through numerous examples that MscalePINN is a necessary extension of  traditional single-scale PINN platforms in situations where
significant multiple scattering effects contribute to determine the effective parameters.
Finally, by exciting with plane waves at different angles, we show that finite-size SHU arrays feature an isotropic homogenized response, i.e. the retrieved effective parameters do not depend on the angle of the incoming radiation. In order to present a comprehensive analysis, we vary the size, shape, incident wavelength, number of scatterers, stealthiness parameter, and direction of excitation for the investigated structures. Our accompanying Supporting Information details all the relevant calculation parameters and provides additional comparisons with single-scale PINN calculations.

\section{\label{sec:methods}Methods}
In optical science and photonics technology, it is often required to solve differential or integro-differential models governing the scattering and transport of vector waves inside complex and heterogeneous materials or in extended media containing resonant optical nanostructures~\cite{tsang_scattering_2000, dal_negro_waves_2022, ishimaru_wave_1978}. 
While many advanced techniques have been developed for the forward solution of such mathematical problems, the multiscale structure of heterogeneous media generally prevents the accurate and efficient solution of inverse scattering problems of relevance to imaging, acoustics, geophysics, remote sensing, and nondestructive testing. Specifically, in the regime of multiple wave scattering where the transport mean free path $\xi_{t}$ is smaller than the system's size $L$, the inversion of differential models becomes a nonlinear and computationally intractable problem for traditional numerical techniques. This prevents the accurate prediction of the desired parameters of multi-particle complex structures from a limited set of available field data, driving the development of alternative and more powerful computational frameworks that leverage automatic learning techniques and optimization methods~\cite{wei_deep-learning_2019, mehta_high-bias_2019, jiang_deep_2021, schmidt_recent_2019, sun_efficient_2018}.

Motivated by this need, recent developments in scientific machine learning (ML) introduced physics-informed neural networks (PINNs) as a viable approach to solving forward and inverse integro-differential problems efficiently and with minimal computational overhead~\cite{lu_deepxde_2021, lu_physics-informed_2021, raissi_physics-informed_2019, chen_physics-informed_2020}. Unlike standard deep learning approaches, PINNs restrict the space of admissible solutions by enforcing the validity of the PDE models governing the actual physics of the problem. This is achieved by using relatively simple feed-forward neural network architectures as trainable surrogate solutions of the partial differential equations (PDEs) on the interior and boundary points of their definition domains and leveraging automatic differentiation (AD) techniques readily available in all the most powerful machine learning packages~\cite{lu_physics-informed_2021,raissi_physics-informed_2019, lu_deepxde_2021,pang_fpinns_2019}. Specifically, PINNs are trained on a set of randomly distributed collocation points to minimize the PDE residues in a suitable norm \cite{lu_physics-informed_2021,raissi_physics-informed_2019}. Therefore, PINNs use only one training dataset to obtain the desired inverse solutions, thus relaxing the burdens often imposed by the massive datasets utilized by alternative, i.e. non-physics-constrained, traditional data-driven deep learning
approaches~\cite{mehta_high-bias_2019}. 
These characteristics render PINNs uniquely effective in solving  differential and integro-differential inverse problems with a minimal overhead compared to the corresponding forward problem~\cite{chen_physics-informed_2020, chen_physics-informed_2022, riganti_auxiliary_2023,sanghvi_embedding_2020}.

\subsection{Mathematical formulation of multiscale PINN}
Recently, it became apparent in the ML community that deep neural networks (DNNs) learn the low-frequency content of available training data quickly and with a good generalization error, but fail to do so when high-frequency data are involved. This general Fourier-type principle creates an implicit spectral bias as DNNs preferentially fit training data using low-frequency functions~\cite{xu_frequency_2020, rahaman_spectral_2019, cheng_wong_learning_2022, ronen_convergence_2019, wang_eigenvector_2021}. To solve this issue in the context of PINNs simulations, the approach of multiscale PINNs (MscalePINNs) was recently introduced that converts the learning and approximation of high-frequency data to that of low-frequency ones, using different sub-networks specialized to learn down-shifted frequency representations of the original datasets and functions~\cite{liu_multi-scale_2020, zhang_correction_2023}. 

To illustrate the approach of the multiscale PINN, we consider a band-limited function $f(\mathbf{x}), \;\mathbf{x}%
\in\mathbb{R}^{d}$, whose Fourier transform $\widehat{f}(\mathbf{k})$ has a
compact support, i.e.,%
\begin{equation}
\text{Supp}\widehat{f}(\mathbf{k})\subset B(K_{\text{max}})=\{\mathbf{k\in
}\mathbb{R}^{d},|\mathbf{k|\leq}K_{\text{max}}\}.\label{support}%
\end{equation}

We can partition the domain $B(K_{\text{max}})$ as an union of $M$ concentric
annulus with uniform or non-uniform width, e.g., for the case of uniform
width $K_{0}$,
\begin{equation}
    \begin{split}
        &A_{i}=\{\mathbf{k\in}\mathbb{R}^{d},(i-1)K_{0}\leq|\mathbf{k|\leq}%
        iK_{0}\},\\
        &K_{0}=K_{\text{max}}/M,\quad1\leq i\leq M, \label{anulnus}
    \end{split}
\end{equation}
so that
\begin{equation}
B(K_{\text{max}})=%
{\displaystyle\bigcup\limits_{i=1}^{M}}
A_{i}.\label{part}%
\end{equation}

As a result, we can decompose the function $\widehat{f}(\mathbf{k})$ in the Fourier domain as before
\begin{equation}
\widehat{f}(\mathbf{k})={\sum\limits_{i=1}^{M}}\chi_{A_{i}}(\mathbf{k}%
)\widehat{f}(\mathbf{k})\triangleq%
{\displaystyle\sum\limits_{i=1}^{M}}
\widehat{f}_{i}(\mathbf{k}),\label{POU}%
\end{equation}
and
\begin{equation}
\text{Supp}\widehat{f}_{i}(\mathbf{k})\subset A_{i}.\label{SupAi}%
\end{equation}

This decomposition in the Fourier space gives a corresponding one in
the physical space%
\begin{equation}
f(\mathbf{x})=
{\displaystyle\sum\limits_{i=1}^{M}}
f_{i}(\mathbf{x}),\label{Partx}%
\end{equation}
where
\begin{equation}
f_{i}(\mathbf{x})=\mathcal{F}^{-1}[\widehat{f}_{i}(\mathbf{k})](\mathbf{x}).\label{convol}%
\end{equation}

From (\ref{SupAi}), we can apply a simple downward scaling to convert the high
frequency region $A_{i}$ to a low-frequency one. Namely, we define a scaled
version of $\widehat{f}_{i}(\mathbf{k})$ as
\begin{equation}
\widehat{f}_{i}^{(\text{scale})}(\mathbf{k})=\widehat{f}_{i}(\alpha
_{i}\mathbf{k}),\qquad\alpha_{i}>1,\label{fkscale}%
\end{equation}
and, correspondingly, in the physical space
\begin{equation}
f_{i}^{(\text{scale})}(\mathbf{x})=\frac{1}{\alpha_{i}^{d}}f_{i}(\frac{1}{\alpha_{i}%
}\mathbf{x}),\label{fscale}%
\end{equation}
or
\begin{equation}
f_{i}(\mathbf{x})=\alpha_{i}^{d}f_{i}^{(\text{scale})}(\alpha_{i}\mathbf{x}).\label{fscale_inv}%
\end{equation}
So, the spectrum of the scaled function $\widehat{f}%
_{i}^{(\text{scale})}(\mathbf{k})$ is of low frequency if $\alpha_{i}$ is chosen large enough,
i.e.,
\begin{equation}
\text{Supp}\widehat{f}_{i}^{(\text{scale})}(\mathbf{k})\subset\{\mathbf{k\in
}\mathbb{R}^{d},\frac{(i-1)K_{0}}{\alpha_{i}}\leq|\mathbf{k|\leq}\frac{iK_{0}%
}{\alpha_{i}}\}.\label{fssup}%
\end{equation}

Now with DNN's preference toward low-frequency learning, with $iK_{0}/\alpha_{i}$ being small, i.e. $iK_{0}/\alpha_{i}=O(2\pi/L)$, where $L$ is the characteristic length of the cluster, we can train a
DNN $f_{\theta^{n_{i}}}(\mathbf{x})$
to learn $f_{i}^{(\text{scale})}(\mathbf{x})$ quickly
\begin{equation}
f_{i}^{(\text{scale})}(\mathbf{x})\sim f_{\theta^{n_{i}}}(\mathbf{x}%
),\label{DNNi}%
\end{equation}
giving an approximation to $f_{i}(\mathbf{x})$ immediately%
\begin{equation}
f_{i}(\mathbf{x})\sim\alpha_{i}^{d}f_{\theta^{n_{i}}}(\alpha_{i}%
\mathbf{x}), \label{fi_app}%
\end{equation}
and,  to $f(\mathbf{x})$ as well
\begin{equation}
f(\mathbf{x})\sim%
{\displaystyle\sum\limits_{i=1}^{M}}
\alpha_{i}^{d}f_{\theta^{n_{i}}}(\alpha_{i}\mathbf{x}),%
\label{eq:f_app}
\end{equation}
giving the format of the MscalePINN \cite{liu_multi-scale_2020, zhang_correction_2023}. And, the scale factors $\alpha_i$ can also be converted into trainable parameters to best fit the target functions. In fact, the factor $\alpha_i^d$ can be absorbed into the weights of the last layer, being linear in most cases,  of the sub-neural network $f_{\theta^{n_{i}}}(\alpha_{i}\mathbf{x})$, the factor $\alpha_i^d$ outside the sub-neural network can be set to be one to avoid involving large values when $\alpha_i$ or $d$ is large without affecting the overall effect of the MscalePINN after training.
Each sub-network with scaled inputs can be written as:

\begin{equation}
    f_\theta(x)=W^{[L-1]}\sigma \circ(\dots(W^{[1]}\sigma\circ(W^{[0]}x+b^{[0]})+b^{[1]})\dots)+b^{[L-1]}
\end{equation}

\begin{figure}
\centering
\includegraphics[scale=0.6]{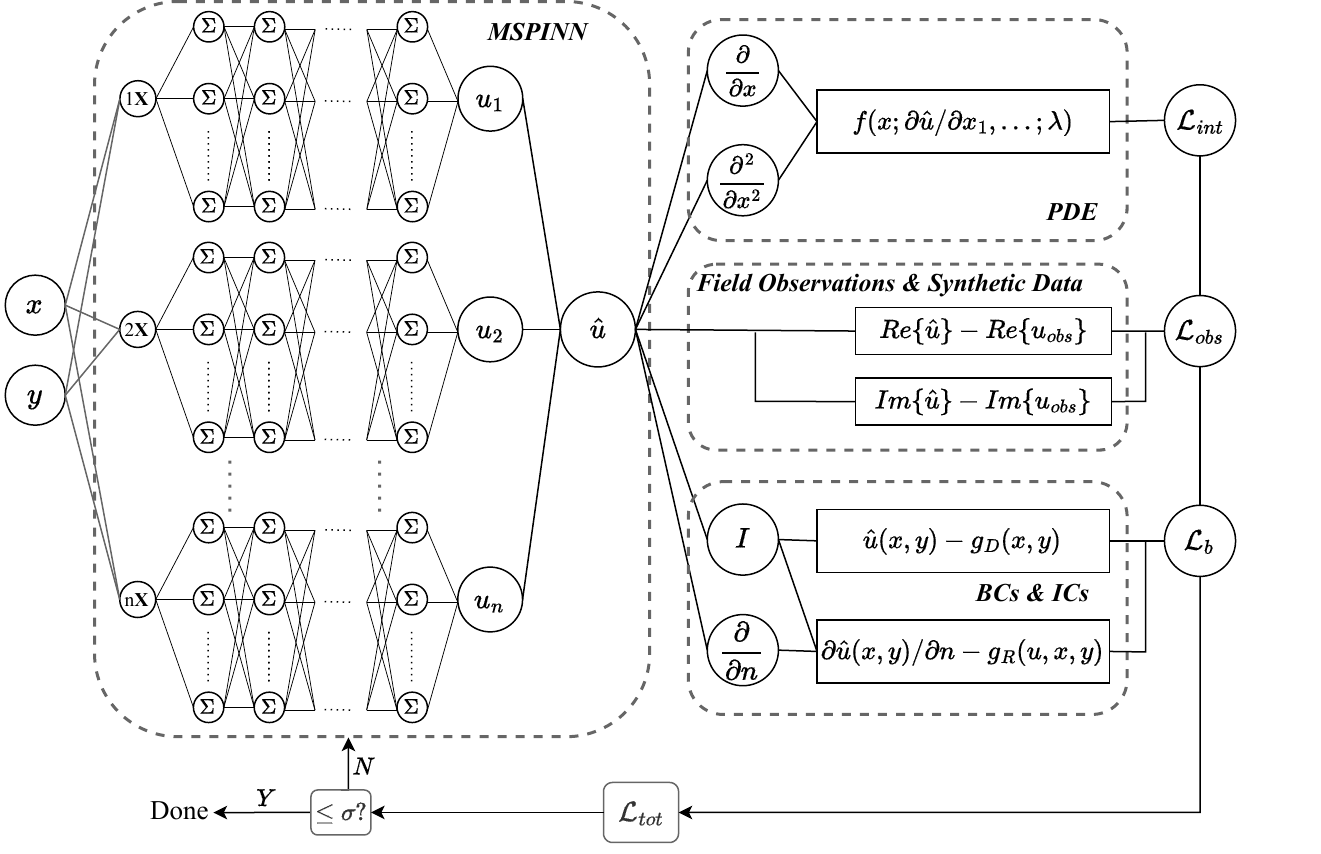}
\caption{Schematics of the multiscale PINN employed for the inverse design of photonic structures}.
\label{fig:MscaleDNN}
\end{figure}

The multiscale network architecture employed in this article is displayed in Fig.~\ref{fig:MscaleDNN}. The spatial input parameters $x,y$ are passed through $n$ independent sub-networks $u_i(\alpha_i x,\alpha_i y;\bm{\tilde{\theta}}_i)$ with different scaling $\alpha_i$ and hyperparameters $\bm{\tilde{\theta}}_i$. The output of each sub-network $u_i$ is then combined into the MscalePINNs solution $\hat{u}(x,y;\bm{\tilde{\theta}})$ and, through automatic differentiation, is then used to satisfy the PDE, boundary, and initial conditions of the differential equation. Specifically, we consider the following PDE problem with the unknown permittivity distribution $\epsilon(x,y;k)$, generally wavevector-dependent, for the surrogate solution $\hat{u}(\bm{x})$ with $\bm{x}=(x_1,\dots,x_d)$ defined on a domain $\Omega \subset \mathbb{R}^d$:
\begin{equation}
    \varphi\left( \bm{x};\hat{u},\frac{\partial\hat{u}}{\partial x_1},\dots,\frac{\partial\hat{u}}{\partial x_d};\frac{\partial^2\hat{u}}{\partial x_1^2},\dots,\frac{\partial^2\hat{u}}{\partial x_1 \partial x_d};\dots; \epsilon \right) = 0
\end{equation}
The calculated values are then combined into the global loss function $\mathcal{L}(\Tilde{\theta})$:
\begin{equation}
    \mathcal{L}(\Tilde{\theta}) = \mathcal{L}_{int}(\Tilde{\theta};\mathcal{N}_{int}) +  \mathcal{L}_{b}(\Tilde{\theta};\mathcal{N}_{b}) + \mathcal{L}_{inv}(\Tilde{\theta};\mathcal{N}_{inv})
\end{equation}
In the loss function above, the component
\begin{equation}
    \begin{split}
        \mathcal{L}_{int}(\bm{\tilde{\theta}};\mathcal{N}_{int})
        =\frac{1}{|\mathcal{N}_{int}|}\sum_{(x,y)\in\mathcal{N}_{int}} \left|\left| \varphi\left( x,y;\hat{u},\frac{\partial\hat{u}}{\partial x},\frac{\partial\hat{u}}{\partial y},\dots, \frac{\partial^2\hat{u}}{\partial y^2}; \epsilon_r \right) \right|\right|^2
    \end{split}
    \label{eq:pde_res}
\end{equation}
represents the loss term calculated for the PDE in the interior of the domain $\Omega$ and
\begin{equation}
    \mathcal{L}_{b}(\bm{\tilde{\theta}};\mathcal{N}_{b}) = \frac{1}{|\mathcal{N}_{b}|}\sum_{(x,y)\in\mathcal{N}_{b}} \left|\left| \mathcal{B}(\hat{u},x,y) \right|\right|^2
\end{equation}
is the loss term for the boundary conditions of the PDE, where  $(x,y)\in \partial\Omega$. Finally, in order to solve general inverse electromagnetic problems, we introduce
\begin{equation}
    \begin{split}    
    \mathcal{L}_{inv}(\bm{\tilde{\theta}};\mathcal{N}_{inv}) =\frac{1}{|\mathcal{N}_{inv}|}\sum_{(x,y)\in\mathcal{N}_{inv}} \left|\left| \text{Re}[\hat{u}(x,y)] - \text{Re}[u_{obs}(x,y)]\right|\right|^2 
    +\left|\left| \text{Im}[\hat{u}(x,y)] - \text{Im}[u_{obs}(x,y)]\right|\right|^2
    \end{split}
\end{equation}
as the inverse loss term calculated on the real and imaginary parts of a complex field obtained through numerical simulations, and $\mathcal{N}_{int},\,\mathcal{N}_{b},\,\mathcal{N}_{inv}$ are the number of residual points for each loss term. 

\subsection{Multiscale PINN for electromagnetic design}
In this article, we deal with electromagnetic parameter retrieval and homogenization problems for which we use the Helmholtz equation to constrain MscalePINNs and retrieve the effective model parameters of optical materials. In particular, we consider the complex Helmholtz equation for inhomogeneous two-dimensional effective media under TM polarization excitation:
\begin{equation}
    \nabla^2 E_z(x,y)+\epsilon_r(x,y;k)k_0^2E_z=0
    \label{eq:helmholtz}
\end{equation}
where $E_z$ is the $z$-component of the electric field, $k_0=\frac{2\pi}{\lambda}$ is the wavenumber in free space, and $\epsilon_r(x,y;k)$ is the relative permittivity of the inhomogeneous effective medium (spatially dependent), which is almost constant in the case of a homogenized effective medium at sufficiently long wavelengths. Because $E_z$ and $\epsilon_r(x,y;k)$ are complex variables, separating Eq.~\ref{eq:helmholtz} into real and imaginary parts yields:
\begin{equation}
    \begin{split}
        \nabla^2\text{Re}[E_z](x,y) =
        -\text{Re}[E_z]\text{Re}[\epsilon_r(x,y;k)]+ \text{Im}[E_z]\text{Im}[\epsilon_r(x,y;k)]k_0^2\\
        \nabla^2\text{Im}[E_z](x,y) =
        -\text{Im}[E_z]\text{Re}[\epsilon_r(x,y;k)]- \text{Re}[E_z]\text{Im}[\epsilon_r(x,y;k)]k_0^2
    \end{split}
    \label{eq:helm_full}
\end{equation}
This framework enables us to predict $\text{Re}[\epsilon_r(x,y;k)]$ and $\text{Im}[\epsilon_r(x,y;k)]$ independently for the inhomogeneous effective medium and therefore to naturally quantify the radiation losses that are particularly difficult to account within the effective index theory~\cite{sihvola_electromagnetic_2008, mishchenko_multiple_2017}.

\begin{figure}
\centering
\includegraphics[scale=1.25]{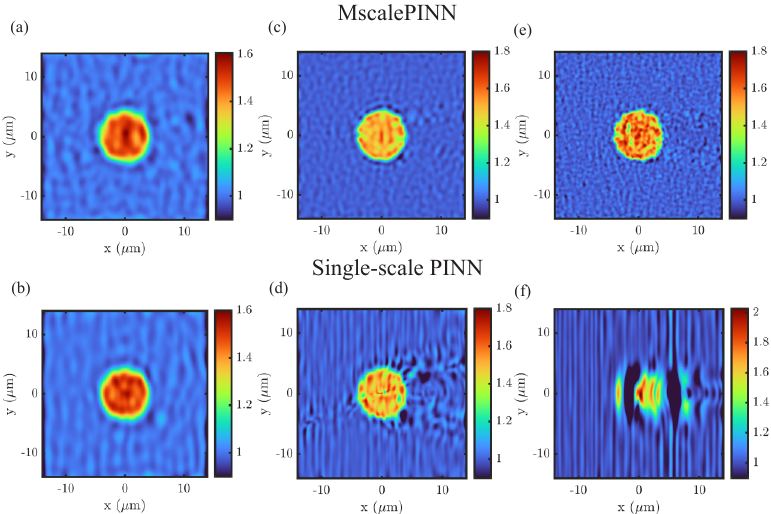}
\caption{(a), (c), and (e) display the accurately retrieved localized effective medium under incoming field $k_0$ vector of $2.0, 4.0,$ and $5.5 \mu$m$^{-1}$ retrieved by MscalePINN. (b), (d), and (f) show that the single-scale PINN can only retrieve an effective medium in the long-wavelength domain, i.e. $k_0=2.0\mu$m$^{-1}$, but fails for the higher $k_0$ vectors $4.0$ and $5.5\mu$m$^{-1}$.}.
\label{fig:com_singleMulti}
\end{figure}

Using the introduced MscalePINNs framework, we inversely retrieved the effective permittivity parameter $\epsilon_r(x,y;k)$ of the Helmholtz equation by training the MscalePINNs over a synthetic dataset composed of complex field values. In particular, we performed finite element method (FEM) simulations of the total complex electric field using COMSOL Multiphysics~\cite{comsol_sftw} and utilized the resulting data as the training set to retrieve $\hat{\epsilon}(x,y;k)$ following the methodology that we introduced in references \cite{chen_physics-informed_2020, chen_physics-informed_2022}. Throughout this paper, we will employ different MscalePINN architectures and, for this reason, we have included Table S1 in the Supporting Information that lists all the relevant hyperparameters employed in our study. All the codes were developed in-house using TensorFlow~\cite{tensorflow2015-whitepaper} and numerical simulations were performed using, depending on the problem, different types of GPUs, i.e., we employed either an NVIDIA P100, NVIDIA V100, or NVIDIA A40.
To establish the accuracy of our results, we perform forward FEM simulations in COMSOL fed by the retrieved $\hat{\epsilon}(x,y;k)$ and we compare real and imaginary parts of the fields with those resulting from considering the actual geometry of the arrays. The resulting errors are quantified by an $L^2$ error norm. If the $L^2$ norm is low enough ($\leq$5\%), then MscalePINN accurately retrieved the effective medium from the original training field. Moreover, for each retrieved $\hat{\epsilon}(x,y;k)$ we calculate the average and standard deviation of its real and imaginary components inside an area with a radius equal to the size of the array. The average over the imaginary part ($\langle\text{Im}[\hat{\epsilon}(x,y;k)]\rangle$) will determine the onset of the transparency regime predicted by Torquato et al.~\cite{torquato_nonlocal_2021, kim_theoretical_2024}. In contrast, the average and standard deviation over the real part ($\langle\text{Re}[\hat{\epsilon}(x,y;k)]\rangle$, $\sigma\lbrace\langle\text{Re}[\hat{\epsilon}(x,y;k)]\rangle$) will characterize the degree of spatial homogeneity of the retrieved effective medium while providing a quantitative metric to compare with results from traditional effective medium theory, such as the Bruggeman mixing formula~\cite{sihvola_electromagnetic_2008}. We will show that, in the long wavelength regime, the MscalePINN results agree with the predictions based on the Bruggeman formula. However, we show that our method can be successfully utilized also at shorter wavelengths and in finite-size device structures of arbitrary shapes beyond the reach of traditional homogenization theories.

\section{\label{sec:results}Results and Discussion}
\subsection{\label{results_0}Multiscale versus single-scale PINN beyond the long-wavelength regime}
We begin our study by showing in Fig.~\ref{fig:com_singleMulti} a direct comparison between the retrieved effective medium profiles obtained using the multiscale and the traditional single-scale PINN  for a stealthy hyperuniform array of $N=663$ dielectric nanocylinders of radius $a=67.5$ nm, packing fraction $\phi=0.16$, and array diameter $L\approx8.5\mu$m. Here $\phi=\rho v(a)$, where $\rho$ is the number density of the array and $v(a)$ is the cross-sectional area of the nanocylinders. For this comparison, we have selected three different datasets for performing the training for the inverse permittivity retrieval. These datasets are the distributions of the complex total fields by the SHU structure at different wavelengths with respect to the predicted "transparency regime" according to the inequality below that characterizes the full extent of the transparency region~\cite{kim_theoretical_2024}:
\begin{equation}
    k_0 \rho^{-1/2} =2\pi\frac{\langle{d}\rangle}{\lambda}\lesssim 1.5, \quad \rho = \frac{N}{\pi R^2}
    \label{eq:inequality}
\end{equation}
where $R$ is the radius of the SHU array. For the SHU structure considered in this example, the inequality (\ref{eq:inequality}) can be satisfied up to the threshold wavenumber $k_{th} \approx 5.4 \mu$m$^{-1}$. Fig.~\ref{fig:com_singleMulti} (a) and (b) show the agreement of the multi-scale and single-scale PINNs for an incident $k_0=2.0\mu$m$^{-1}$ (corresponding to $\lambda=3.14\mu$m). This agreement is expected since this $k_{0}$ value is well below the threshold value for the transparency region. In Fig.~\ref{fig:com_singleMulti} (c) and (d), however, we notice that the single-scale PINN architecture cannot properly retrieve an effective medium for $k_0=4.0\mu$m$^{-1}$ ($\lambda=1.57\mu$m) since this wavenumber value is much closer to the edge of the transparency region. In contrast, the MscalePINN has no trouble retrieving a well-localized and homogeneous effective medium even at $k_0=4.0\mu$m$^{-1}$. Finally, in panels (e) and (f), we show the complete failure of single-scale PINN to retrieve a physically meaningful effective medium distribution when the wavenumber of the incoming radiation crosses the transparency edge and enters the strong multiple scattering regime at $k_{th} = 5.5 \mu$m$^{-1}$ ($\lambda=1.14$). On the other hand, the MscalePINN retrieves a well-localized, albeit inhomogeneous, effective medium also in this case. Therefore, the developed MscalePINN is a powerful extension of the traditional single-scale PINN that becomes necessary when solving the inverse parameter retrieval problem in the multiple scattering regime. Additional characterizations of the investigated SHU structure and the utilized training fields can be found in Figures S2 and S3 of our Supporting Information. In the next section, we will address the role of disorder averaging in retrieving the effective medium of  stealthy hyperuniform and uncorrelated Poisson arrays.

\begin{table}[t!]
\centering
\caption{Comparison of the real part of the homogenized effective permittivity $\hat{\epsilon}(x,y;k)$ between the ensemble average and single realization for SHU and Poisson structures with the same size.} 
\begin{tabular}{lrrr}
Structure & Calculation & $\langle$Re$[\hat{\epsilon}(x,y;k)]\rangle$ & $\sigma\lbrace\langle\text{Re}[\hat{\epsilon}(x,y;k)]\rangle$ \\
\midrule
\multirow{2}{*}{SHU} & Single realization & 1.391 & 2.8\%  \\
 & Ensemble average & 1.404 & 1.9\%  \\
 \midrule
\multirow{2}{*}{Poisson} & Single realization  & 1.412 & 8.8\%  \\
 & Ensemble average & 1.413 & 3.4\%  \\
\bottomrule
\end{tabular}
\label{table:comp}
\end{table}

\subsection{\label{results_1}MscalePINN analysis of SHU ensemble average}
\begin{figure}
\centering
\includegraphics[scale=1.5]{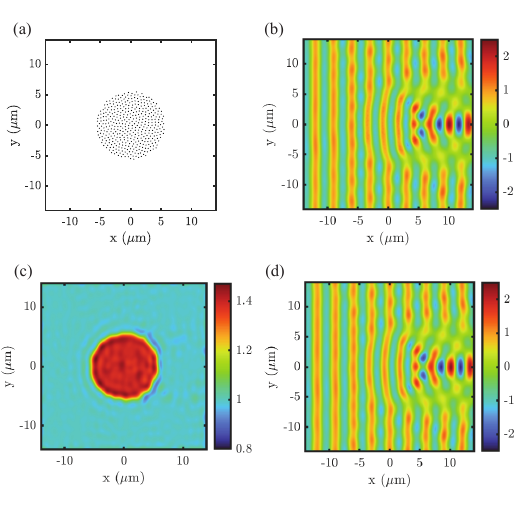}
\caption{(a) SHU array of 396 particles, $\chi=0.5, \epsilon_r=3.0$, $\langle d \rangle / \lambda = 0.15$ employed for the MscalePINNs homogenization validation. (b) The real part of the FEM electric field inverse train dataset is employed to train MscalePINNs to homogenize panel (a). The incident plane wave wavelength is $\lambda=3.0\mu$m. (c) MscalePINN's predicted relative permittivity distribution, which is then used to perform a forward COMSOL simulation displayed in panel (d) to compare with the ``true" field used in training.}
\label{fig:validation}
\end{figure}

We are interested here in retrieving the homogenized complex permittivity of the SHU array with diameter $L\approx10\mu$m composed of $N=396$ dielectric nanocylinders of radius $a=125$nm, displayed in Fig.~\ref{fig:validation}(a). The corresponding structure factor $S(\bm{k})$ is shown in Figure S1 of the Supporting Information. For this example, we used $\chi=0.5, \epsilon_r=3.0,$  $\langle d \rangle / \lambda = 0.15$, and $\phi=0.2$, where $\langle d \rangle$ is the average first-neighbor distance of the cylinders in the array. The MscalePINNs utilized to solve this inverse problem is a 4-scale MscalePINNs with 4 layer, each with 64 neurons, and it is trained on the real and imaginary parts of the total electric field considering the excitation wavelength  $\lambda=3.0\mu$m. 

In Fig.~\ref{fig:validation}(b) we display the real part of the electric field distribution used during training. The effective permittivity profile $\hat{\epsilon}(x,y;k)$ retrieved by MscalePINNs for a single realization of the investigated SHU array is displayed in panel (c). Remarkably, the effective permittivity is well-localized within the geometrical support of the array with a spatially uniform distribution quantified by the average value  $\langle$Re$[\hat{\epsilon}(x,y;k)]\rangle=1.39\pm2.8\%$. To better characterize the retrieved homogeneous permittivity profile in the static regime we also calculated the effective medium theory prediction using the Bruggeman mixing formula valid for the bulk case~\cite{sihvola_electromagnetic_2008}:

\begin{equation}
    \sum_i f_i \frac{\epsilon_i-\epsilon_e}{\epsilon_i+\epsilon_e} = 0
    \label{eq:Bruggeman}
\end{equation}
where $\epsilon_e$ is the effective permittivity, $f_i$ is the filling fraction, and $\epsilon_i$ is the permittivity of the $i$-th component. For a two-phase system under TM polarized incident radiation, Eq.~\ref{eq:Bruggeman} reduces to~\cite{kim_theoretical_2024}:
\begin{equation}
    \epsilon_{brugg} = f_1 \epsilon_1 + f_2\epsilon_2
\end{equation}
For the system studied in Fig.~\ref{fig:validation}(c) it yields $\epsilon_{brugg}=1.4$, differing only by 0.6\% from MscalePINNs and within the uncertainty range predicted. In Fig.~\ref{fig:validation}(d) we display the real part of the total electric field obtained via a forward FEM calculation performed using the retrieved permittivity profile from panel (c), which is then used to calculate the $L^2$ error including both the real and imaginary part of the training field. The obtained complex field error in the retrieval of the permittivity parameter was found to be $1.1\%$, demonstrating the high accuracy of the solution achieved by the developed MscalePINNs.

To further investigate the quality of the reconstruction, we also performed an ensemble average of 10 different SHU configurations all generated with the same stealthiness parameter $\chi=0.5, \epsilon_r=3.0,$ and $N=395\pm 5$ and with constant $\phi=0.2$. Our findings are summarized in the first row of Table~\ref{table:comp} which displays almost identical results to the single realization case when point-wise spatial averaging is performed over the SHU realizations. As a comparison, we also show in Table~\ref{table:comp} the results of the same ensemble averaging procedure performed over 10 different realizations of Poisson uncorrelated random (UR) structures with $\epsilon_r=3.0,$ $N=395\pm 5$, and $\phi=0.2$. We note that, compared to the SHU configuration, the single realization for the Poisson structure features a significant inhomogeneity in the spatial distribution of Re$[\hat{\epsilon}(x,y;k)]$ due to the presence of larger fluctuations among the different disorder realizations. However, even in this case the retrieved effective medium permittivity for the Poisson point pattern has an $L^2$ error lower than 5\% when the forward total FEM field of the array was compared to the one obtained from the inversely retrieved permittivity. This indicated that MscalePINNs has retrieved an accurate spatially dependent permittivity, i.e., an inhomogeneous effective medium,  demonstrating the importance of the more general network approach developed here. Therefore, we conclude that MscalePINNs retrieved an accurate field distribution and $\hat{\epsilon}(x,y;k)$ for both the SHU and Poisson arrays and that, at the single realization level, a homogeneous permittivity can only be retrieved for the SHU structures. In the next section, we investigate the behavior of the MscalePINN at shorter wavelengths for both the SHU structures and the uncorrelated Poisson arrays focusing on a scattering regime where $\hat{\epsilon}(x,y;k)$ cannot be homogenized using conventional mixing formulas~\cite{sihvola_electromagnetic_2008}.

\subsection{SHU and Poisson point patterns beyond the long wavelength regime}
\begin{figure}
\centering
\includegraphics[scale=1.5]{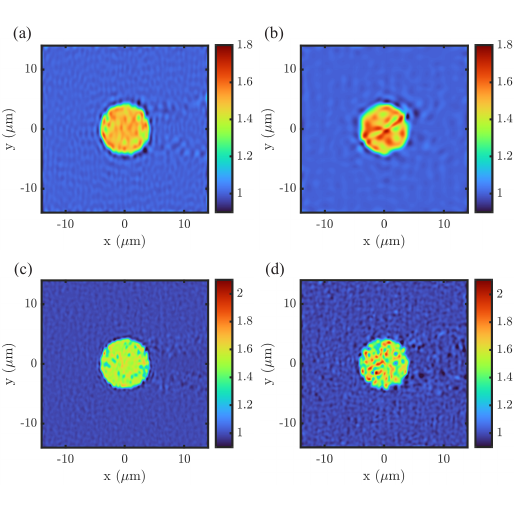}
\caption{(a) and (b) Comparison between the retrieved permittivity profile $\text{Re}[\hat{\epsilon}(x,y;k)]$ of a stealthy hyperuniform array with $\chi=0.3$, $N=663$ particles and the inhomogeneous effective medium of a Poisson array of $N=661$ particles. The inverse total field used in training had $\lambda=2.09\mu$m ($\langle d \rangle / \lambda = 0.1$), or $k_0 = 3.0\mu$m$^{-1}$. (c) and (d) Comparison of the same structures used in panels (a) and (b) but with a lower wavelength of the incident scattering field, $\lambda=1.26\mu$m ($\langle d \rangle / \lambda = 0.14$), or $k_0 = 5.0\mu$m$^{-1}$. Already at $\lambda=2.09\mu$m, the Poisson array displays evident losses and a more inhomogeneous retrieved $\text{Re}[\hat{\epsilon}(x,y;k)]$.}
\label{fig:comparison}
\end{figure}

To compare the effective medium behavior of SHU and Poisson structures we generate a Poisson array comparable to the SHU structure discussed in Section~\ref{results_0} with $N\approx660$ dielectric nanocylinders of radius $a=67.5$ nm, $\phi=0.16$, and diameter $L\approx8.5\mu$m. Their structure factors $S(\bm{k})$ are shown in Figure S3 of the Supporting Information. Fig.~\ref{fig:comparison}(a) and (c) show the retrieved $\hat{\epsilon}(x,y;k)$ through the MscalePINN for $k_0 = 3.0\mu$m$^{-1}$ and 5.0$\mu$m$^{-1}$, respectively (corresponding to $\lambda=2.09\mu$m and $1.26\mu$m). The considered SHU structure has $\chi=0.3$ and $\epsilon_r=4.0$. We observe that
in panels (a) and (c) both solutions are more spatially homogeneous than the corresponding ones shown in panels (b) and (d) for the Poisson pattern. Specifically, for the SHU structure we found $\langle\text{Re}[\hat{\epsilon}(x,y;k)]\rangle=1.48\pm4\%$ and $\langle\text{Im}[\hat{\epsilon}(x,y;k)]\rangle=10^{-3}\approx 0$ for $\lambda=2.09\mu$m and $\langle\text{Re}[\hat{\epsilon}(x,y;k)]\rangle=1.51\pm6.0 \%$ and $\langle\text{Im}[\hat{\epsilon}(x,y;k)]\rangle=10^{-3}\approx 0$ for $\lambda=1.26\mu$m. Moreover, we note that for the short-wavelength simulation near the edge of the transparency region (i.e., $k_{0} = 5.5\mu$m$^{-1}$), the MscalePINN retrieves an average value of $\text{Re}[\hat{\epsilon}(x,y;k)]$ that is different from the Bruggeman predicted value of 1.48. 

However, for the retrieved permittivity of the Poisson structures shown in panels (b) and (d), the degree of spatial non-uniformity becomes apparent already away from the SHU critical value of $5.5\mu$m$^{-1}$. Fig.~\ref{fig:comparison}(b) shows the inhomogeneous profile of the real part for the Poisson's effective permittivity profile $\hat{\epsilon}(x,y;k)$ predicted by the MscalePINNs for $\lambda=2.09\mu$m. For this inhomogeneous effective medium, $\langle\text{Re}[\hat{\epsilon}(x,y;k)]\rangle=1.49\pm6.8\%$, but MscalePINNs failed to retrieve the imaginary component of $\hat{\epsilon}(x,y;k)$, despite an $L^2$ error on the FEM validation of 4\%. The inability to retrieve a homogenized effective permittivity is even more apparent at shorter wavelengths (panel d), where for $\lambda=1.26\mu$m the MscalePINN correctly predicts an effective permittivity profile $\hat{\epsilon}(x,y;k)$ with $\langle\text{Re}[\hat{\epsilon}(x,y;k)]\rangle=1.52\pm11\%$ shown in Fig.~\ref{fig:comparison}(d), but fails to predict the imaginary component of $\hat{\epsilon}(x,y;k)$. This evident failure to homogenize the random Poisson pattern for $\lambda=2.09\mu$m and $1.26\mu$m compared to the SHU structure is both qualitative and quantitative. In fact, the spatial non-uniformity of Re$[\hat{\epsilon}(x,y;k)]$ for the Poisson structure, measured by the standard deviation $\sigma$, is consistently greater than that of the SHU structure for the same incoming wavelength. We have included an additional figure that summarizes these findings in Figure S5 of the Supporting Information where we show the consistent difference in spatial non-uniformity for the Poisson patterns, quantified by both $\sigma$ and $\langle\text{Im}[\hat{\epsilon}(x,y;k)]\rangle$.
In Figure S5, we have included the MscalePINN's prediction on the single realization beyond the $k_0$ critical value in the grey-shaded region, where the $L^2$ error was higher with a value of 24\%. We also note that for the SHU array, the real part of $\hat{\epsilon}(x,y;k)$ becomes less homogeneous as we approach the edge of the predicted transparency region. In this case, the  MscalePINN continues to accurately retrieve an effective medium until the predicted edge of $k_0=5.5\mu$m$^{-1}$, while it fails for the Poisson arrays. This scenario supports the conclusion that SHU structures can be more easily homogenized than traditional Poisson random media and that their homogenization eventually fails at larger incident wave vectors than their Poisson counterparts. In the next section, we will address the effects of size scaling on the MscalePINN predictions of the $\hat{\epsilon}(x,y;k)$ retrieved permittivity distribution.

\begin{figure}
\centering
\includegraphics[scale=1.5]{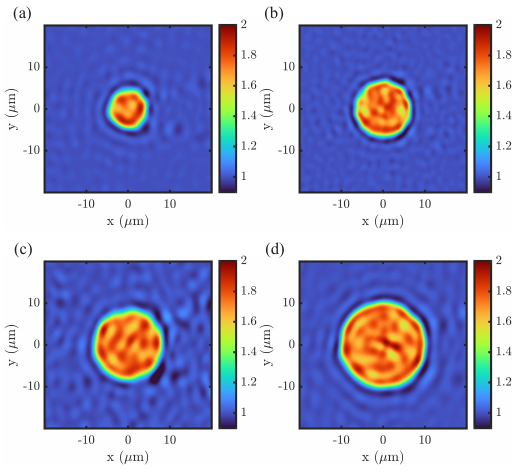}
\caption{(a)-(d) Scaling analysis of the retrieved effective medium performed on four SHU arrays with 299, 633, 1002, and 1553 particles, respectively, and $\langle d \rangle / \lambda = 0.06$. The average of MscalePINNs’s permittivity profile displayed in the four panels all have a computed average value within the region of  $\langle\text{Re}[\hat{\epsilon}(x,y;k)]\rangle=1.70\pm5.9\%$, independently of the size of the hyperuniform array.}
\label{fig:scaling}
\end{figure}

\subsection{\label{sec:transparency}Transparency of finite-size SHU structures}
Recent work by Torquato and Kim \cite{torquato_nonlocal_2021,kim_theoretical_2024} led to an exact non-local strong-contrast expansion of the effective dynamic dielectric tensor  $\epsilon (x,y)$ in the thermodynamic limit~\cite{torquato_nonlocal_2021}, and more recently they extended these results beyond the long-wavelength regime for layered and transversely isotropic media~\cite{kim_theoretical_2024}. Their work provides an analytical prediction for the wavelength range in which SHU structures achieve perfect transparency or, equivalently, for the wavelength regime where the effective dielectric constant has a zero imaginary part. However, to the best of our knowledge, no previous work has established if this transparency prediction is modified by finite-size arrays. In order to address this open question we performed a study over several SHU arrays with different sizes and numbers of pillars $N=299, 633, 1002,$ and $1553$. We also computed the structure factors for the corresponding point patterns and displayed the results in Fig. S6 of the Supporting Information. Interestingly, we note that characteristic stealthy hyperuniform behavior begins to manifest itself already at relatively small $N$. 

In Fig.~\ref{fig:scaling} we display the retrieved effective permittivity $\hat{\epsilon}(x,y;k)$ for these four stealthy hyperuniform structures with parameters $\chi=0.3, \epsilon_r=4.0,$ and $\phi=0.25$, which were kept constant for all the arrays to compare with the theoretical predictions for the infinite bulk limit shown in Ref.~\cite{kim_theoretical_2024}. We train a 4-scale MscalePINNs with 2 layers and 64 neurons each using FEM computed forward fields at plane wave excitation wavelength $\lambda=6.28\mu$m, corresponding to the regime of perfect transparency predicted by Torquato et al.~\cite{kim_theoretical_2024}. We display the real part of the retrieved permittivity profiles $\hat{\epsilon}(x,y;k)$ in order of increasing array size in Fig.~\ref{fig:scaling}(a)-(d), and compute the mean and standard deviation inside of the array region. The average of the retrieved effective dielectric function for the four structures is $\langle\text{Re}[\hat{\epsilon}(x,y;k)]\rangle=1.7\pm5.9\%$ and $\langle\text{Im}[\hat{\epsilon}(x,y;k)]\rangle=10^{-4}$, independent of the array size. This result is extremely close to the predicted value in Ref.~\cite{kim_theoretical_2024} for transversely isotropic media with $\chi=0.3, \epsilon_r=4.0,$ and $\phi=0.25$. Furthermore, due to the relatively large wavelength regime, the predicted $\hat{\epsilon}$ agrees with the Bruggeman mixing formula of $\epsilon_{brugg}=1.75$ with a 3\% error. Therefore, from this analysis, we conclude that finite-size scaling does not perturb appreciably the value of the retrieved dielectric constant of SHU arrays. In the next sections, we conclude our analysis by first demonstrating that MscalePINN can be employed to retrieve homogenized effective media that leverage the isotropic homogenized response featured by SHU structures, a property that is highly desired for engineering angle-insensitive effective media of finite size. Furthermore, we show a relevant application to photonics inverse design, specifically in the context of wave guiding and focusing structures. Lastly, we discuss how it is possible to employ MscalePINN for the inverse design of binary optical metamaterials. 

\begin{figure}
\centering
\includegraphics[scale=1.5]{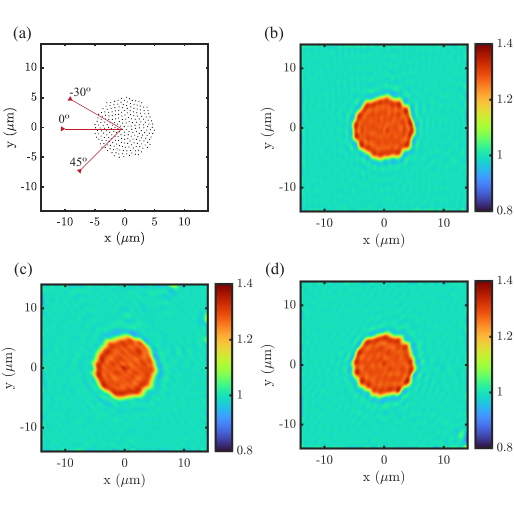}
\caption{(a) Schematics of the angles employed to study the angular independence in the homogenization of a SHU array of $N=236$ particles, $\chi=0.5$, $\epsilon_r=3.0$,  $\langle d \rangle / \lambda = 0.09$, and packing fraction $\phi=0.20$. (b)-(d) MscalePINNs’s reconstruction of the homogenized permittivity profile $\hat{\epsilon}(x,y;k)$ retrieved by training on incident light with wavelength $\lambda=6.28\mu$m at three different angles: $\theta=0^\circ,45^\circ,-30^\circ$. The averages of  $\langle\text{Re}[\hat{\epsilon}(x,y;k)]\rangle$ in the homogenized regions agree and show that the homogenization is independent of the angle of incidence.}
\label{fig:angle_chi50}
\end{figure}

\subsection{Isotropic response of finite-size SHU structures}
The development of complex photonic media with isotropic scattering responses is important to for engineering device applications as it naturally result in robust performances with enhanced light-matter coupling.
In order to demonstrate numerically that homogenized stealthy hyperuniform arrays are isotropic with respect to the direction of the incoming electric field, we perform angle-dependent simulations on a stealthy hyperuniform array $N=236$ with $\chi=0.5$, $\epsilon_r=3.0$, and $\phi=0.20$. In Fig.~\ref{fig:angle_chi50}(a) we display the SHU array with the angles employed to generate the forward FEM numerical simulations utilized to train the 4-scale MscalePINNs with 4 layers by 64 neurons each. As in the previous studies, we have included the corresponding structure factor $S(\bm{k})$ in the Supporting Information Figure S7, together with the inverse training FEM fields displayed in Figure S8. The predicted homogenized permittivity profiles $\hat{\epsilon}(x,y;k)$ are shown in Fig.~\ref{fig:angle_chi50}(b)-(d), displaying the MscalePINNs precision in capturing contour features on the boundary of the hyperuniform array. All three permittivity profiles present an accurate agreement in the real and imaginary part of $\hat{\epsilon}(x,y;k)$, with $\langle\text{Re}[\hat{\epsilon}(x,y;k)]\rangle=1.28\pm3\%$ and $\langle\text{Im}[\hat{\epsilon}(x,y;k)]\rangle=10^{-4}$. To confirm that the three homogenized effective media have indeed the same electromagnetic response, we selected the homogenized permittivity profile $\hat{\epsilon}(x,y;k)$ trained with the incoming radiation at $\theta=0^\circ$ and performed two forward FEM simulations at $\theta=45^\circ$ and $-30^\circ$. We then computed an $L^2$ error between the forward field used in training on the SHU array at $\theta=45^\circ$ and $-30^\circ$ and the one just recomputed by utilizing the homogenized structure trained with the incoming radiation at $\theta=0^\circ$. The two $L^2$ errors were $1.2\%$ and $2.4\%$, respectively, showing that the effective medium retrieved by MscalePINNs when trained with incoming radiation at $\theta=0^\circ$ reproduced the same training FEM field when the incoming angle was set to $\theta=45^\circ$ and $-30^\circ$. In conclusion, we showed numerically that single-realization finite-size SHU arrays feature an isotropic homogenized response to the incoming radiation. 

\begin{figure}
\centering
\includegraphics[scale=1.5]{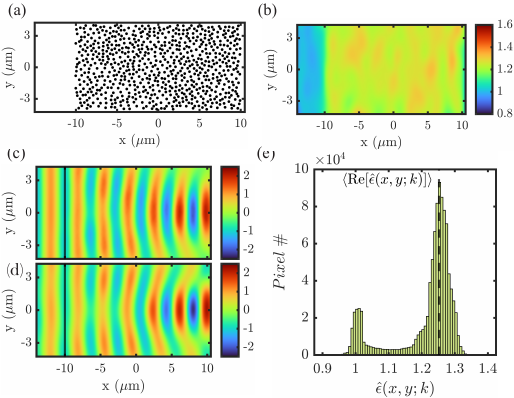}
\caption{(a) Geometry employed for the forward FEM field simulation displayed in (c)  with $k_0=1.57\mu$m$^{-1}$, where the blank space on the left is intended for wave propagation outside of the structure. (b) Inversely retrieved $\text{Re}[\hat{\epsilon}(x,y;k)]$ with its distribution displayed in (e), where we have labeled the average value $\langle\text{Re}[\hat{\epsilon}(x,y;k)]\rangle=1.252$ inside the effective medium. The distribution of values is bimodal around the effective medium average and air outside of the structure. (d) The forward FEM field computed through the inversely retrieved structure in (b).}
\label{fig:waveguide}
\end{figure}

\subsection{MscalePINN for the design of a SHU waveguide}
There has been growing interest in exploiting the transparency region of SHU structures due to their robustness against multiple scattering. In fact, recent work has shown that this transparency regime can be leveraged to improve the design of photonic materials and devices such as waveguides~\cite{milosevic_hyperuniform_2019, cheron_wave_2022, romero-garcia_wave_2021}. 
Here, we employ MscalePINN to demonstrate the homogenization of a stealthy hyperuniform photonic medium in a rectangular waveguide geometry similar to the one originally investigated in  Ref.~\cite{cheron_wave_2022} to demonstrate enhanced wave transport in beyond the diffusive regime. Specifically, we consider the rectangular stealthy hyperuniform distribution of $N\approx700$ cylindrical scatterers with $\epsilon_r=2.2$ inside a waveguide with stealthiness $\chi=30\%$ shown in Fig.~\ref{fig:waveguide} (a). This stealthiness value is within the range discussed by Cheron et al. We then performed a forward FEM simulation and trained the MscalePINN on the inverse training dataset field shown in Fig.~\ref{fig:waveguide} (c) with $k_0=1.57\mu$m$^{-1}$, which is away from the predicted transparency threshold wavenumber $k_{th}\approx3.0\mu$m$^{-1}$. The homogenized retrieved effective medium $\text{Re}[\hat{\epsilon}(x,y;k)]$ is shown in panel (b), and its distribution of values is displayed in the histogram in panel (e). We found that the average value for the real part of the permittivity was $\langle\text{Re}[\hat{\epsilon}(x,y;k)]\rangle=1.252\pm3\%$, and the $L^2$ error between the total field obtained on the homogeneous $\hat{\epsilon}(x,y;k)$, displayed in panel (d), and the training field was 6\%. These results show that MscalePINN can be employed to accurately retrieve a homogenized effective medium in a waveguide configuration leveraging the enhanced transparency properties of SHU arrays of scattering nanocylinders.  This enables the efficient design of materials and devices with desired permittivity distributions and enhanced optical transparency beyond the diffusion limit of traditional random media, a problem of great relevance for photonics applications~\cite{cheron_wave_2022}.
This includes the ability to design spatially inhomogeneous and compact metamaterials that implement low-loss mode transformation and control of the effective index of refraction for on-chip integrated photonic couplers in the near-infrared spectral range \cite{cheben_subwavelength_2018}. In the next and final section, we describe an implementation of the MscalePINN architecture that enables the inverse design of optical materials with a choice of binary permittivity values.

\subsection{MscalePINN for the inverse design of binary optical materials}
Throughout this paper, we have shown that MscalePINN is a powerful extension of traditional single-scale PINN in the context of photonic inverse design from the long- to short-wavelength regime. In particular, using $k_{th}$ as a metric, we displayed the MscalePINN's ability to retrieve effective media for wavelengths close to the multiple scattering regime. However, the retrieved structures so far allowed for a continuously varying permittivity, a feature that makes their fabrication very challenging. In order to overcome this limitation, we introduce here a different MscalePINN architecture that enables the accurate inverse design of binary optical materials and metamaterials with an optimal permittivity distribution $\hat{\epsilon}(x,y;k)$ to achieve a functionality. Importantly, in this implementation the binary values of the retrieved permittivity profile can be largely determined by the designer, thus making our approach compatible with a large class of available optical materials.

So far, we have employed the multiscale architecture displayed in Fig.~\ref{fig:MscaleDNN}, where the final layer's output $\hat{u}$ is a linear combination of the subnetworks' outputs $u_i$. As a result, the effective medium featured a continuous distribution of $\hat{\epsilon}(x,y;k)$ values, only limited by the PDE and synthetic data characteristics. However, in most engineering applications the effective medium design is constrained by the limitations of available materials, i.e., we must choose a well-characterized material with a suitable refractive index. When considering the inverse design of the optimal shape of a functional optical structure in air, given the availability of a chosen optical material, we must modify the MscalePINN to enable retrieving a spatial distribution $\hat{\epsilon}(x,y;k)$ with binary values corresponding to the one of air and of the chosen material. We achieve this task by bounding the network's final output with a custom sigmoid activation function:
\begin{equation}
    \sigma_{cust} = \frac{\gamma}{1+e^{-\xi*x}}+\eta
    \label{eq:cust_sig}
\end{equation}
Where $\gamma, \xi, \eta$ are user-defined parameters determining the material and background effective permittivity of the retrieved effective medium. In particular, $\eta$ is the background medium, usually set to 1.0 (air), while $\gamma$ determines the epsilon values above 1.0. For example, an effective medium of relative permittivity 2.2 immersed in air should have $\gamma=1.2$ and $\eta=1.0$. Finally, $\xi$ is the slope of the final layer's sigmoid function, determining how sharp the binarization in $\hat{\epsilon}$  should be during training. This is the hardest parameter to set, as higher values of $\xi$ do ensure sharper transitions, but MscalePINN becomes difficult to train due to very large gradients during the back-propagation of the network. These large gradients are caused by the large derivatives of the sigmoid function that approximates a sharp step function. In practice, we experienced that it is safer to keep $\xi$ at the default value of 1.0 during training, and in the examples that follow, we will describe an alternative way to improve binarization while keeping the training stable. Furthermore, in the following discussion, we will consider two different scenarios, namely the retrieval of a constant-epsilon medium (the homogenization case) and a binary-epsilon medium embedded in an air background (the multiple-scattering case). We refer to the latter case as to the multiple-scattering one because we consider plane wave excitations close to the given structure's $k_{th}$ value. The MscalePINN architectures employed hereafter are similar, as they both use a multiscale network such as the one displayed in Fig.~\ref{fig:MscaleDNN} for the real and imaginary part of the electric field. Still, they differ in the coupled $\hat{\epsilon}$ network that retrieves the permittivity profile. For the homogenization architecture, we employ a full-domain single-scale network with a binary sigmoid output. Instead, for the multiple-scattering case, we employ a multiscale network with a binary output trained only in a square region of interest. In practice, what this means is that in the homogenization case, we let MscalePINN choose where the transition between air and the material occurs in the computational domain $\Omega$, while in the multiple-scattering case we let MscalePINN determine the transition between the two materials only within a region of interest while keeping $\hat{\epsilon}$ constant at unit values outside.
\begin{figure}
\centering
\includegraphics[scale=1.25]{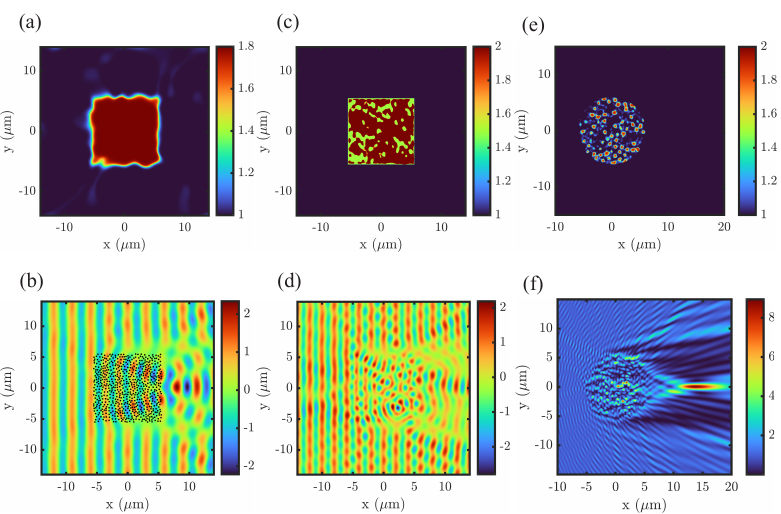}
\caption{(a) Binary structure retrieved on the $\lambda=3\mu$m field using the modified MscalePINN approach with the custom sigmoid activation function from Eq.~\ref{eq:cust_sig}. (b) Forward simulation on the structure displayed in (a) with the original SHU pattern superimposed. The error between the forward field in (b) and the original was 4\%. (c) Two-level binarized structure retrieved at $\lambda=2.0\mu$m, or $k_0=3.14\mu$m$^{-1}$, on the SHU pattern shown in (b). Panel (d) displays the forward field through the binarized structure displayed in (c), which had an $L^2$ error of 3\% on the original. (e) Focusing effective medium retrieved by MscalePINN at $\lambda=1.0\mu$m with $\hat{\epsilon}=1.9$. (f)
For better visualization, we display the focusing intensity of the total field obtained by performing a forward FEM simulation through the effective medium displayed in panel (e).}
\label{fig:binary}
\end{figure}

In Fig.~\ref{fig:binary} (a) we display an example of MscalePINN binarization for a square SHU structure under plane wave excitation at $\lambda=3.0\mu$m. For this example, we have picked the same structural parameters from the arrays in Fig.~\ref{fig:com_singleMulti}, but we cut the array in a square instead of a circle to display the versatility of MscalePINN when retrieving objects with sharp corners. In Fig.~\ref{fig:binary} (b), we have overlayed the SHU array to the field retrieved with a forward FEM simulation on the homogeneous structure. In this simulation, we chose to find an effective medium with $\hat{\epsilon}=1.8$, so we set $\gamma=0.8,\xi=1.0,\eta=1.0$. As the results show, the effective medium is binarized, with an error on the true field of 4\%.
In Fig.~\ref{fig:binary} (c) we display the binary effective medium retrieved when the square structure employed in panel (a) is excited by a plane wave at $\lambda=2.0\mu$m. For this wavelength, it is impossible to retrieve a single homogeneous medium because we are at the edge of the transparency region, with $k_0=3.14$ and $k_{th}=3.5$, so we restrict the training of the $\hat{\epsilon}$ MscalePINN to a square subset $(x_{sq},y_{sq})$ of $\Omega$ in the central region with dimensions $11\mu\text{m}\times11\mu\text{m}$. To display the power of MscalePINN, we set $\gamma=0.5,\xi=1.0,\eta=1.5$, which are meant to model a binary material with $\hat{\epsilon}_1=1.5$ and $\hat{\epsilon}_2=2.0$. To ensure a proper binarization, we add an additional loss term $\mathcal{L}_{bin}$ defined as:
\begin{equation}
    \mathcal{L}_{bin}(\bm{\tilde{\theta}};\mathcal{N}_{bin}) = \frac{1}{|\mathcal{N}_{bin}|}\sum_{(x_{sq},y_{sq})\in\mathcal{N}_{bin}} \left|\left| 
    \Theta(\hat{\epsilon}(x_{sq},y_{sq};k))-\hat{\epsilon}(x_{sq},y_{sq};k)
    \right|\right|^2
    \label{eq:Lbin}
\end{equation}
Here, $\Theta$ is a step function centered at $\eta + \gamma/2$. We switch on $\mathcal{L}_{bin}$ after $1/4$ of the total training epochs to ensure that the $\hat{\epsilon}$ MscalePINN retrieves is more binarized. In panel (c) we display the binarized structure obtained on post-processing the final output of MscalePINN with the $\Theta$ function described above. The pre-processed output is displayed in section S9 of the Supporting Information. This is a common strategy employed in optimization algorithms that are constrained by fabrication requirements, and it ensures that the final output of MscalePINN is very close to a fabricable structure~\cite{christiansen_inverse_2021}. The $L^2$ error for both the structure in panel (c) and the pre-processed output is close to 3\%, where we have displayed the forward field on the binarized one in panel (d). 
Finally, we inversely retrieve the structure of a binary material from training on the focusing fields studied in Ref.~\cite{zhu_inverse_2023}, with $\lambda=1.0\mu$m. The training datasets corresponding to the focusing field are shown in section S10 of the Supporting Information. Panel (f) displays the intensity of the total field obtained with a forward simulation on the structure retrieved in panel (e), demonstrating a clear focusing behavior at a distance of $15\mu$m. For this simulation, we chose $\gamma=0.9,\xi=1.0,\eta=1.0$, corresponding to pillars with $\hat{\epsilon}=1.9$.  Despite the wavelength being in the multiple-scattering regime, the retrieved structure accurately reproduced the focusing field with an $L^2$ error below 8\%.

\section{Conclusions}
In this article, we developed and applied the multiscale physics-informed neural network framework to inversely retrieve the effective dielectric permittivity of finite-size arrays of scattering nanocylinders with stealthy hyperuniform and uncorrelated Poisson geometries. Through numerous examples, we established that MscalePINN is a necessary powerful extension of traditional single-scale PINN architectures when dealing with multiple scattering contributions in the retrieval of the effective medium parameters of complex media, thus enabling a systematic methodology to retrieve the general spatial dependence of the effective dielectric behavior of scattering arrays in device-relevant geometries beyond traditional homogenization theories. In particular, we demonstrated the existence of a transparency region in finite-size SHU structures beyond the long-wavelength approximation, enabling effective and isotropic homogenization even without disorder-averaging, in contrast to the case of uncorrelated Poisson random patterns.
Specifically, we found that the retrieved permittivity distribution $\hat{\epsilon}(x,y;k)$ obtained for a single-realization of stealthy hyperuniform disorder agrees with the ensemble average calculations with an error close to 1\%, whereas a large standard deviation of $\hat{\epsilon}(x,y;k)$ is obtained for Poisson arrays of comparable sizes. 
Importantly, we showed that homogenized SHU structures feature isotropic responses to an incoming plane wave excitation that are not appreciably modified by their finite size, which is a highly desired characteristic for the engineering of angle-insensitive photonic media and metamaterial devices. Furthermore, we applied our approach to the accurate homogenization of a rectangular waveguide structure starting from a scattering SHU array with enhanced transparency.
Lastly, we show how to achieve the inverse design of a binary permittivity distributions trained on customized field profiles using a modified MscalePINN architecture, thus enabling the efficient design of functional photonic structures in the multiple scattering regime. Our work provides an efficient route towards the discovery of novel structures with effective medium properties arising from the interaction of disordered scattering geometries and vector or scalar waves of arbitrary nature, including acoustic, mechanical, and quantum wave phenomena.

\section{Supporting Information}
The Supporting Information presents details of the investigated structures and simulations performed, additional comparisons with Poisson patterns, and the MscalePINN hyperparameters used in training. 
\medskip

\section{Acknowledgments}
L. D. N. acknowledges the support from the U.S. Army Research Office, RF-Center managed by Dr. T. Oder (Grant \#W911NF-22-2-0158). S. T. acknowledges the support of the U.S. Army Research Office under Cooperative Agreement No. W911NF-22-2-0103. The work of W. Cai is supported by the US National Science Foundation grant DMS-2207449.

\bibliographystyle{unsrt}  
\bibliography{SHU_v4}

\end{document}